\documentclass[traditabstract]{aa}
\usepackage{amsmath}
\usepackage{txfonts}
\usepackage[switch]{lineno}

\usepackage{graphicx}

\usepackage{url}
\usepackage{framed}
\usepackage{natbib,ifthen}

\usepackage[table,usenames,dvipsnames]{xcolor}
\definecolor{linkcolor}{rgb}{0.6,0,0}
\definecolor{citecolor}{rgb}{0,0,0.75}
\definecolor{urlcolor}{rgb}{0.12,0.46,0.7}
\usepackage[breaklinks, colorlinks, urlcolor=urlcolor, linkcolor=linkcolor,citecolor=citecolor,pdfencoding=auto]{hyperref}

\usepackage{longtable}
\usepackage{multirow}

\def\setsymbol#1#2{\expandafter\def\csname #1\endcsname{#2}}
\def\getsymbol#1{\csname #1\endcsname}

\def\Planck{\textit{Planck}}

\newbox\tablebox    \newdimen\tablewidth
\def\leaderfil{\leaders\hbox to 5pt{\hss.\hss}\hfil}
\def\endPlancktable{\tablewidth=\columnwidth 
    $$\hss\copy\tablebox\hss$$
    \vskip-\lastskip\vskip -2pt}

\def\tablenote#1 #2\par{\begingroup \parindent=0.8em
    \abovedisplayshortskip=0pt\belowdisplayshortskip=0pt
    \noindent
    $$\hss\vbox{\hsize\tablewidth \hangindent=\parindent \hangafter=1 \noindent
    \hbox to \parindent{$^#1$\hss}\strut#2\strut\par}\hss$$
    \endgroup}
\def\doubleline{\vskip 3pt\hrule \vskip 1.5pt \hrule \vskip 5pt}

\def\L2{\ifmmode L_2\else $L_2$\fi}

\def\DeltaT{\ifmmode \Delta T\else $\Delta T$\fi}
\def\deltat{\ifmmode \Delta t\else $\Delta t$\fi}
\def\fknee{\ifmmode f_{\rm knee}\else $f_{\rm knee}$\fi}
\def\Fmax{\ifmmode F_{\rm max}\else $F_{\rm max}$\fi}
\def\solar{\ifmmode{\rm M}_{\mathord\odot}\else${\rm M}_{\mathord\odot}$\fi}
\def\Msolar{\ifmmode{\rm M}_{\mathord\odot}\else${\rm M}_{\mathord\odot}$\fi}
\def\Lsolar{\ifmmode{\rm L}_{\mathord\odot}\else${\rm L}_{\mathord\odot}$\fi}
\def\inv{\ifmmode^{-1}\else$^{-1}$\fi}
\def\mo{\ifmmode^{-1}\else$^{-1}$\fi}
\def\sup#1{\ifmmode ^{\rm #1}\else $^{\rm #1}$\fi}
\def\expo#1{\ifmmode \times 10^{#1}\else $\times 10^{#1}$\fi}
\def\,{\thinspace}
\def\lsim{\mathrel{\raise .4ex\hbox{\rlap{$<$}\lower 1.2ex\hbox{$\sim$}}}}
\def\gsim{\mathrel{\raise .4ex\hbox{\rlap{$>$}\lower 1.2ex\hbox{$\sim$}}}}

\def\simprop{\mathrel{\raise .4ex\hbox{\rlap{$\propto$}\lower 1.2ex\hbox{$\sim$}}}}
\def\deg{\ifmmode^\circ\else$^\circ$\fi}
\def\pdeg{\ifmmode $\setbox0=\hbox{$^{\circ}$}\rlap{\hskip.11\wd0 .}$^{\circ}
          \else \setbox0=\hbox{$^{\circ}$}\rlap{\hskip.11\wd0 .}$^{\circ}$\fi}
\def\arcs{\ifmmode {^{\scriptstyle\prime\prime}}
          \else $^{\scriptstyle\prime\prime}$\fi}
\def\arcm{\ifmmode {^{\scriptstyle\prime}}
          \else $^{\scriptstyle\prime}$\fi}
\newdimen\sa  \newdimen\sb
\def\parcs{\sa=.07em \sb=.03em
     \ifmmode \hbox{\rlap{.}}^{\scriptstyle\prime\kern -\sb\prime}\hbox{\kern -\sa}
     \else \rlap{.}$^{\scriptstyle\prime\kern -\sb\prime}$\kern -\sa\fi}
\def\parcm{\sa=.08em \sb=.03em
     \ifmmode \hbox{\rlap{.}\kern\sa}^{\scriptstyle\prime}\hbox{\kern-\sb}
     \else \rlap{.}\kern\sa$^{\scriptstyle\prime}$\kern-\sb\fi}
\def\ra[#1 #2 #3.#4]{#1\sup{h}#2\sup{m}#3\sup{s}\llap.#4}
\def\dec[#1 #2 #3.#4]{#1\deg#2\arcm#3\arcs\llap.#4}
\def\deco[#1 #2 #3]{#1\deg#2\arcm#3\arcs}
\def\rra[#1 #2]{#1\sup{h}#2\sup{m}}

\def\dots{\relax\ifmmode \ldots\else $\ldots$\fi}
\def\WHzsr{\ifmmode $W\,Hz\mo\,sr\mo$\else W\,Hz\mo\,sr\mo\fi}
\def\mHz{\ifmmode $\,mHz$\else \,mHz\fi}
\def\GHz{\ifmmode $\,GHz$\else \,GHz\fi}
\def\mKs{\ifmmode $\,mK\,s$^{1/2}\else \,mK\,s$^{1/2}$\fi}
\def\muKs{\ifmmode \,\mu$K\,s$^{1/2}\else \,$\mu$K\,s$^{1/2}$\fi}
\def\muKRJs{\ifmmode \,\mu$K$_{\rm RJ}$\,s$^{1/2}\else \,$\mu$K$_{\rm RJ}$\,s$^{1/2}$\fi}
\def\muKHz{\ifmmode \,\mu$K\,Hz$^{-1/2}\else \,$\mu$K\,Hz$^{-1/2}$\fi}
\def\MJysr{\ifmmode \,$MJy\,sr\mo$\else \,MJy\,sr\mo\fi}
\def\MJysrmK{\ifmmode \,$MJy\,sr\mo$\,mK$_{\rm CMB}\mo\else \,MJy\,sr\mo\,mK$_{\rm CMB}\mo$\fi}
\def\microns{\ifmmode \,\mu$m$\else \,$\mu$m\fi}

\def\muK{\ifmmode \,\mu$K$\else \,$\mu$\hbox{K}\fi}
\def\microK{\ifmmode \,\mu$K$\else \,$\mu$\hbox{K}\fi}
\def\muW{\ifmmode \,\mu$W$\else \,$\mu$\hbox{W}\fi}
\def\kms{\ifmmode $\,km\,s$^{-1}\else \,km\,s$^{-1}$\fi}
\def\kmsMpc{\ifmmode $\,\kms\,Mpc\mo$\else \,\kms\,Mpc\mo\fi}
\providecommand{\sorthelp}[1]{}

\providecommand{\sorthelp}[1]{}

\bibpunct{(}{)}{;}{a}{}{,} 

\newcommand{\fnurl}[1]{\footnote{\url{#1}}}

\newcommand{\artdeco}{\texttt{artDeco}}
\newcommand{\Artdeco}{\texttt{ArtDeco}}
\newcommand{\Madam}{\texttt{Madam}}

\newcommand{\nside}{\ifmmode N_\mathrm{side}
                                       \else $N_\mathrm{side}$\fi}
\newcommand{\healpix}{\texttt{HEALPix}}

\newcommand{\lmax}{\ensuremath{\ell_{\text{max}}}}
\newcommand{\kmax}{\ensuremath{k_{\text{max}}}}

\newcommand{\aslm}{$a_{s\ell m}$}

\begin{document}

\title{Beam-deconvolved \Planck\ LFI maps}

\author
{E.~Keih\"anen\inst{1},
V.~Lindholm\inst{1},
M. Lopez-Caniego\inst{3},
M.~Maris\inst{4},
M.~Reinecke\inst{2},
M.~Sandri\inst{5},
 and
A.-S.~Suur-Uski\inst{1} }

\authorrunning{Keih\"anen et al.}

\institute{
University of Helsinki, Department of Physics,
P.O.~Box 64, FIN-00014, Helsinki, Finland \\
\email{elina.keihanen@helsinki.fi}
\and
Max-Planck-Institut f\"ur Astrophysik, Karl-Schwarzschild-Str.~1, 85741 Garching, Germany
\and
European Space Agency, ESAC, Planck Science Office,
Camino bajo del Castillo, s/n, Urbanizaci\'on Villafranca del Castillo,
Villanueva de la Ca\~nada, Madrid, Spain
\and
INAF - Osservatorio Astronomico di Trieste, Via G.B. Tiepolo 11, Trieste, Italy
\and
INAF - OAS Bologna, Istituto Nazionale di Astrofisica - Osservatorio di Astrofisica e Scienza dello Spazio di Bologna, 
Area della Ricerca del CNR, Via Gobetti 101, 40129, Bologna, Italy
}

\abstract{
The \Planck\ Collaboration made its final data release in 2018.
In this paper we describe beam-deconvolution map products made from \Planck\  Low Frequency Instrument (LFI) data
using the \artdeco\ deconvolution code to symmetrize the effective beam.
The deconvolution results are auxiliary data products, available through the Planck Legacy Archive.
Analysis of these deconvolved survey difference maps reveals signs of residual signal in the 30-GHz and 44-GHz 
frequency channels. We produce low-resolution maps and corresponding noise covariance matrices (NCVMs).
The NCVMs agree reasonably well with the half-ring noise estimates except for 44\,GHz, where we observe an asymmetry
between $EE$ and $BB$ noise spectra, possibly a sign of further unresolved systematic error.
{In contrast to the official \Planck\ LFI maps, the beam-deconvolution maps have not been corrected for bandpass mismatch,
and the residual noise is not well approximated by white noise.}
}


\maketitle

\section{Introduction}

In 2018, the \Planck
\footnote{Planck ({\tt http://www.esa.int/Planck}) is a project of the European Space Agency (ESA) with instruments provided by two scientific consortia funded by ESA member states 
(in particular the lead countries France and Italy), with contributions from NASA (USA) and telescope reflectors provided by a collaboration between ESA and a scientific 
consortium led and funded by Denmark.} 
collaboration made its third major data release, PR3 \citep{planck2016-l01}, based on the entire mission, which covers 48 months for the Low Frequency Instrument (LFI)
and 29 months for the High Frequency Instrument (HFI).
The release comprises an extensive set of maps, including frequency maps for the full mission, partial maps for 12- and 6-month periods, and maps of  various detector combinations.
Descriptions of the data processing pipelines from time-ordered data to final map products
are given in \cite{planck2016-l02} and \cite{planck2016-l03}.

The PR3 maps are not corrected for beam shape; each data sample is assigned entirely to the pixel where the centre of the beam falls.  The beam properties of \Planck\ LFI are described in \cite{planck2014-a05}.  Beam shapes affect the maps in several ways.  An effect immediately visible by eye is that the image of a point source is not symmetric, but is deformed according to the shape of the detector beam.  More importantly, from a cosmology point of view, the usual map-making procedure produces maps with an effective beam that varies across the sky.   The {\tt FEBeCop} code \citep{mitra2010} constructs an effective beam, which describes the local deformation taking into account the detector beam shape and the local distribution of measurements.  The effective beam is different for every sky pixel, and also depends on the survey and detector combination.

Beam asymmetries complicate the analysis of survey difference maps and other null maps,
which play an important role in data validation. Two maps made of the same local sky region
 at different times are not identical, since the pattern of beam orientations is different
 due to the scanning strategy. 
Furthermore, beam shape mismatch is a source of leakage of temperature signals to polarization.
Detectors with different beam shapes collect different signals from the same point on the sky.
The difference is then falsely interpreted as a polarization signature by the usual map-making methods.
The LFI applies horn-uniform weighting \citep{planck2016-l02} to alleviate the problem.
With horn-uniform weighting, the leakage depends mainly on the mismatch between beams
within a horn, which is usually smaller than the mismatch between horns.

At the power spectrum level, the average beam smearing is described by a beam window function.
A conventional beam window does not correct for leakage effects,
but a more sophisticated matrix window formalism \citep{planck2014-a05} provides 
a tool for the correction of the latter effect as well.

In this work we perform beam deconvolution on LFI data.
We use the \artdeco\ deconvolution code \citep{keihanen2012} 
to correct for the asymmetric smearing produced by the beams of the  detectors.  
We take as our starting point the calibrated and destriped timelines for LFI detectors,
now publicly available through the Planck Legacy Archive (PLA) \footnote{http://pla.esac.esa.int/pla}.
The input data are thus already calibrated, and cleaned of correlated noise, as far as possible.

\Artdeco\ operates in harmonic space, and yields a harmonic representation 
of the sky signal without beam smoothing.
This can be converted into a conventional sky map,
but because the harmonic representation is limited up to a cut-off \lmax\ that
depends on the beam width,
we must smooth the maps with a Gaussian window
to eliminate ringing artefacts.
A point source in a deconvolved map is thus not reduced to a point,
but into a symmetric Gaussian shape of finite width.  

Deconvolution products offer several benefits over the usual un-deconvolved maps.
Deconvolution map-making produces a map with a symmetric effective beam
with the same shape at every point on the sky, regardless of detector or survey combination.
In our case the effective beam has a simple Gaussian form,
which is easy to handle in further processing steps.
In harmonic space the situation is simpler still, since there is no smoothing
involved.
As potential use cases that could benefit from deconvolved inputs, we mention component separation and analysis of point sources.

Deconvolution map-making eliminates the leakage of temperature
signal to polarization via beam shape mismatch.
Another source of leakage is the mismatch in frequency response between detectors.
This is not affected by deconvolution, and thus it is still present in the deconvolved maps.
With beam effects out of the way, we can perform a more robust analysis
of survey difference maps, possibly revealing residual systematics that were 
previously hidden by beam effects.

In cosmological analysis it is useful to analyse the lowest multipoles of the cosmic microwave background anisotropy spectrum separately.
Deconvolution offers a natural way of extracting the low-multipole signal,
as it operates primarily in harmonic space.
With this in mind,  we construct pixelized low-resolution maps
and corresponding noise covariance matrices (NCVMs).
To validate the NCVMs, we compare them to half-ring noise estimates. 

It is outside the scope of this paper to perform full-scale analysis of the deconvolved maps up to power 
spectra and cosmological parameters.  However, the deconvolved maps
and related NCVM are available through PLA for the community to study.

Beam symmetrization does not come without a price.
Deconvolution alters the residual noise in a non-trivial way.
In the PR3 maps, the residual noise at high multipoles is 
approximatively white. In the deconvolved maps this is no longer
true, since the deconvolution process creates noise correlations
between neighbouring pixels. 
At low multipoles, the effects are captured in the NCVMs.
At high multipoles we rely on half-ring noise estimates. 

This paper is organised as follows. 
In Sect. \ref{sec:deconvolution} we review the deconvolution process.
In Sect. \ref{sec:highresolution} we describe the deconvolved high-resolution \Planck\ LFI maps,
and highlight differences with respect to 
the PR3 maps. We also analyse survey-difference maps and 
search for signs of residual systematics.
In Sect. \ref{sec:lowresolution}  we describe the production of deconvolved low-resolution maps
and their NCVMs.  We validate the NCVM products both in harmonic and pixel space.
Finally, we give some conclusions in Sect. \ref{sec:conclusions}.


\section{Deconvolution of Planck LFI  data}
\label{sec:deconvolution}

\subsection{ArtDeco}

The \artdeco\ beam deconvolver takes as input the time-ordered information (TOI) and 
detector pointings for one or several detectors, and their beam shapes.
The code produces as output a set of harmonic coefficients \aslm\ of the sky,
up to some maximum value \lmax. 
The algorithm is described in \cite{keihanen2012}.

The beam shapes are given in the form of the harmonic expansion of the beam shape, $b_{s\ell k}$,
where $s=0,\pm2$.
Index $s=0$ corresponds to the temperature signal, while $s=\pm2$ components are related to polarization.
Parameter \kmax\ defines the maximum absolute value of index $k$,
and controls the level of beam asymmetry taken into account in the analysis.
{In this analysis we account for beam asymmetry up to \kmax=6.
The simulations carried out in \citet{keihanen2012}
show that this is sufficient in the case of elliptic beams.
The real \Planck\ beams have more complex structure, although
the true ellipticities are smaller than those assumed in the earlier simulations.
In another context \citep{keihanen2016} we have used the realistic LFI beams
for a power spectrum analysis involving beam deconvolution,
and demonstrated that the \kmax=6 limit captures well the structure
of the realistic LFI beams.
}

From the harmonic coefficients \aslm, one can further construct a beam-deconvolved sky map,
through a spherical harmonic transform.
The harmonic coefficients must be smoothed before the transform.
The smoothing serves two purposes:  it eliminates ringing artefacts that arise
from the sharp cut-off of the spectrum at \lmax;
and it suppresses the small-scale noise, which deconvolution tends to amplify.
The smoothing width that fulfills these two requirements is comparable to 
or slightly larger than the width of the actual detector beam.

\subsection{Input TOI}
\label{sec:inputTOI}

The \artdeco\ deconvolver operates under the assumption that noise in the input data is
uncorrelated between samples.
The raw \Planck-LFI TOI do not fulfill this requirement, since the data stream is contaminated by
 correlated $1/f$ noise.
The \Planck\ LFI pipeline uses the
\Madam\ destriper \citep{keihanen2010} for noise removal and mapmaking.
Pre-whitened TOIs are available in the form of ring objects,
where samples on the same pointing period and with the same position are coadded
to make one ring element.  The input data are thus already calibrated and cleaned of correlated noise.
We use these ring objects as inputs for the deconvolution operation.
These ring objects are available through the PLA. 

Even after destriping, a small amount of residual correlated noise is still present;
this adds to the signal power in the lowest multipoles.
The properties of residual noise
and its effect on deconvolution are addressed in \cite{keihanen2015}.
Deconvolution amplifies the noise in the high-multipole regime of the angular power spectrum,
and creates correlation between neighbouring pixels.
At low multipoles, the main effect is the rescaling of the signal (including noise)
in compensation for the missing beam power.

When deconvolving partial data sets, for instance single-year data or detector subsets,
there is some freedom in combining the destriping and deconvolution steps.
One option is to use the full frequency data set for destriping, to maximally utilise all data available and
 to remove the correlated noise as accurately as possible,
and then use a subset of the cleaned TOI for deconvolution. This is referred to as the full destriping option.
Another option is to use the same data subset for both the destriping and the deconvolution steps,
which is called independent destriping.
The latter option leaves considerably more residual noise,
but offers the benefit that the residual noise is uncorrelated between data sets.

The input ring objects available through the PLA are produced by destriping the full frequency data set.
Since we take these objects as inputs, we are automatically using the full destriping option.
We thus have for each frequency one set of ring objects,
which serves as input in all deconvolution runs for that frequency.

\subsection{Beams}

\Planck\ LFI beams consist of three components \citep{planck2014-a05}.
The main beam is defined as extending to 1\pdeg9, 1\pdeg3, and 0\pdeg9 from the beam boresight
at 30, 44, and 70\,GHz, respectively.
The near sidelobes cover the region between the main beam and $5^\circ$.
The remaining part is defined as far sidelobes.
An estimate of the sidelobe contribution to the signal is already removed as part of the calibration step  \citep{planck2014-a03}.
We consider in the deconvolution process the main and near sidelobe beam components. 

We use the radiometer scanning beams as described in  \cite{planck2014-a05}.
The scanning beams include the effective beam elongation caused by the satellite motion. 
These must be distinguished  from the effective beams that take into account the scanning strategy and
are a superposition of different beam orientations.  The effective beam is different for every survey and detector combination.
Characteristics of both beam types for LFI, {together with plots of the beam shapes,} can be found in {\cite{planck2013-p02d} and in} \cite{planck2014-a05}.

{Effective beams are a combination of intrinsic detector beam shape and pixel shape.
They are typically less asymmetric (measured by fitted ellipticity) than the scanning beams, due to the averaging over beam orientations.}
The difference is particularly striking at 44\,GHz; the ellipticities of radiometer scanning beams vary between 
1.188 (LFI25S) and 1.388 (LFI24M) \citep{planck2014-a05}
while the ellipticity of the combined effective beam for the whole mission is only 1.035.
Effective beams and scanning beams reflect different aspects of the same beam.
For instance, the image of a point source in an un-deconvolved map resembles the image of the effective beam.
On the other hand, temperature leakage to polarization via beam mismatch is dependent on the mismatch
between scanning beams, since the polarization signal is obtained effectively by differencing the timelines 
of two radiometers.

\subsection{Deconvolution products}

We have produced deconvolved maps for all LFI frequencies,  single horns, and for horn pairs with
complementary polarization directions. 
For each horn combination, we produce a full mission map, single year maps,
and individual single survey maps.   
A single survey refers to a period of approximately six months;
exact definitions of the surveys can be found in \cite{planck2014-a03}.
Deconvolution parameters for different detector combinations are listed in 
Table~\ref{tab:deconvolution_products}.
The same parameters apply to all survey combinations.


\begin{table}[ht!]
\caption{Deconvolution parameters by horn combination, deconvolution parameters \lmax\ and \kmax, polarization (T/F), 
and FWHM (Full-width half-maximum) width of the smoothing kernel.  The same deconvolution parameters apply to all surveys.}
\label{tab:deconvolution_products}
\vskip -8mm
\setbox\tablebox=\vbox{
\newdimen\digitwidth
\setbox0=\hbox{\rm 0}
\digitwidth=\wd0
\catcode`*=\active
\def*{\kern\digitwidth}
\newdimen\signwidth
\setbox0=\hbox{+}
\signwidth=\wd0
\catcode`!=\active
\def!{\kern\signwidth}
\newdimen\decimalwidth
\setbox0=\hbox{.}
\decimalwidth=\wd0
\catcode`@=\active
\def@{\kern\signwidth}
\halign{\hbox to 0.9in{#\leaderfil}\tabskip=2.0em&
    \hfil#\hfil&
    \hfil#\hfil&
    \hfil#\hfil&
    \hfil#\hfil\tabskip=0em\cr
\noalign{\doubleline}
\omit\hfil Horns\hfil&\omit\hfil \lmax\hfil&\omit\hfil \kmax\hfil&Pol&FWHM [\arcm]\cr 
\noalign{\vskip 5pt\hrule\vskip 5pt}
 \noalign{\vskip 4pt}
\omit{\bf 30\,GHz}\hfil\cr
\noalign{\vskip 4pt}
\hglue 1em 27--28& 800& 6&  T& 40\cr
\hglue 1em 27& 800& 6&  T&  40\cr
\hglue 1em 28& 800& 6&  T&  40\cr
\noalign{\vskip 5pt}
\omit{\bf 44\,GHz}\hfil\cr
\noalign{\vskip 4pt}
\hglue 1em 24--26& 1000& 6&  T&   30\cr
\hglue 1em 25/26& 800& 6&  T&   40\cr
\hglue 1em 24& 1000& 6&  F&   30\cr
\noalign{\vskip 5pt}
\omit{\bf 70\,GHz}\hfil\cr
\noalign{\vskip 4pt}
\hglue 1em 18--23& 1500& 6&  T&  20\cr
\hglue 1em 18/23& 1500& 6&  T&  20\cr
\hglue 1em 19/22& 1500& 6&  T&  20\cr
\hglue 1em 20/21& 1500& 6&  T&  20\cr
}}
\endPlancktable
\end{table}

For each survey and detector combination, we release both the
harmonic coefficients, which are the primary deconvolution output,
and a pixelized \healpix\footnote{https://healpix.jpl.nasa.gov} map \citep{gorski2005}  constructed from them.

The pixelized maps have resolution \nside=1024, corresponding to $3\parcm4\times3\parcm4$ pixels.
We apply smoothing by a symmetric Gaussian beam 
of FWHM (full-width half-maximum) width $40'$ (30\,GHz), $30'$ (44\,GHz) or $20'$ (70\,GHz).
These widths were chosen to suppress the angular power of the map below a fraction of $10^{-6}$ of the unsuppressed value
at the cut-off \lmax.  This is sufficient to safely remove all visible ringing around the strongest sources.

The smoothing window is

\begin{equation}
W(\ell) = \begin{cases}
\exp[-\sigma^2\ell(\ell+1)] \quad  & \hbox{(temperature)} \\
\exp[-\sigma^2(\ell(\ell+1)-4)] \quad &  \hbox{(polarization),} \\
\end{cases}
\end{equation}
where the $\sigma$ parameter of the Gaussian function is related to the FWHM
through $\sigma$=FWHM$/\sqrt{8\ln(2)}$,
that is $\sigma=16\parcm9864$, $12\parcm7398$, $8\parcm4932$ for FWHM=$40'$, $30'$, $20'$, respectively.
The effective beam in pixel space is thus a Gaussian, with width as given above.
Since we release the deconvolved harmonic coefficients along with the maps,
an interested user can easily produce alternative maps with any desired smoothing
width.


\section{High-resolution maps}
\label{sec:highresolution}

\subsection{Full-mission temperature maps}

\begin{figure*}
\includegraphics[width=18cm]{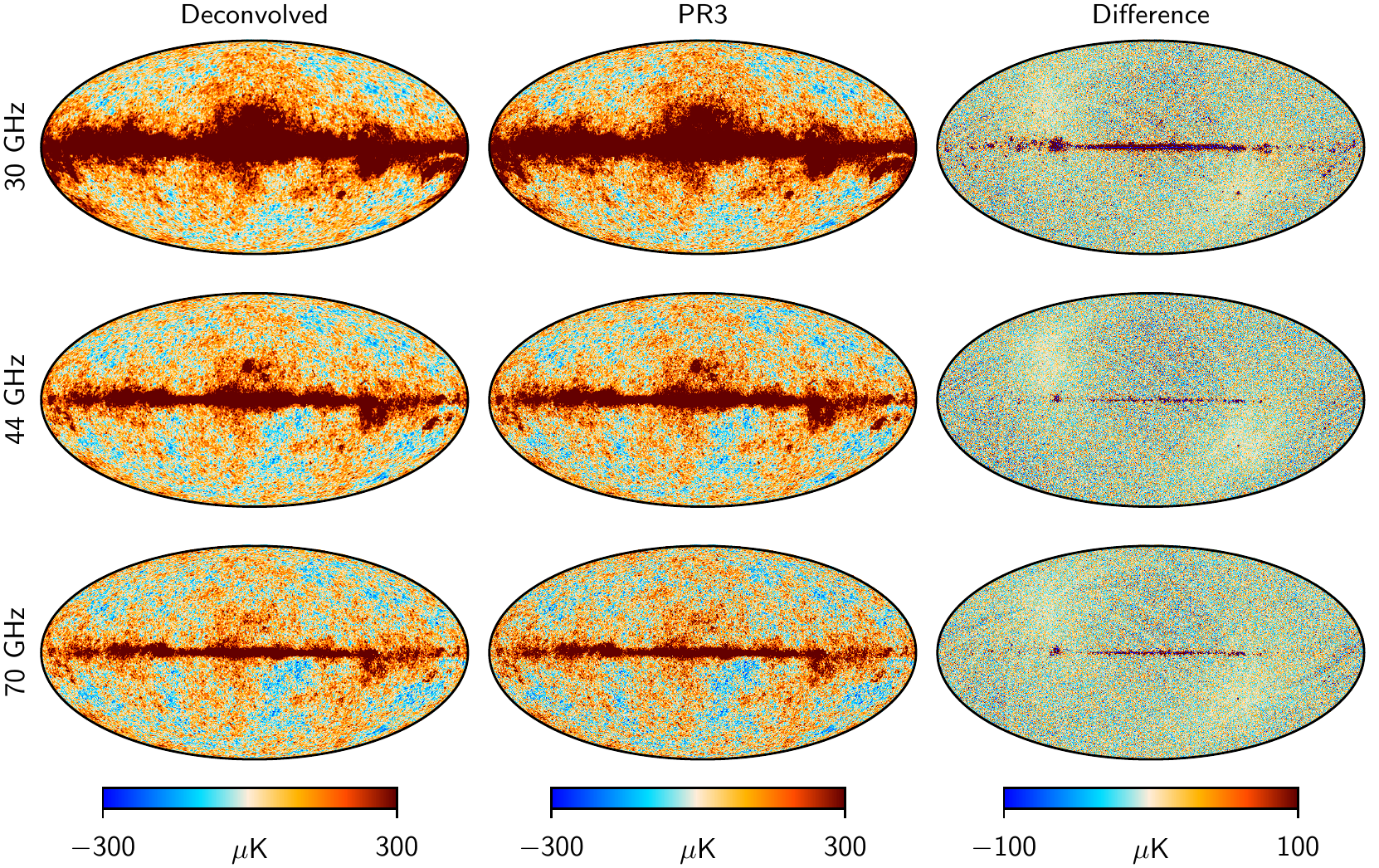}
\caption{From left to right, the deconvolved temperature maps for 30, 44, and 70\,GHz, the corresponding PR3 maps, and the differences between the two.  The data include four years of observations for all radiometers of the respective frequency channel.
The deconvolved maps are smoothed with a Gaussian beam of FWHM=$40'$ (30\,GHz),
$30'$ (44\,GHz), or $20'$ (70\,GHz).
We subtracted from the deconvolved maps the same monopole that had been subtracted 
from the  PR3 map.
The difference maps are dominated by white noise, which is suppressed in the deconvolved maps.
The temperature scales of the maps are $\pm300\,\mu$K for temperature and $\pm100\,\mu$K
for the difference.
}
\label{fig:frequency_maps}
\end{figure*}

We examine first the deconvolved, full-mission temperature maps for 30, 44, and 70\,GHz.
Deconvolved maps and the PR3 products are shown side by side in 
Fig.~\ref{fig:frequency_maps}.
The two sets of maps are visually very similar.
Variations become more visible when we look at the difference map (deconvolved$-$PR3),
shown in the third column.


The difference {outside the Galactic region} is dominated by noise.
The smoothing applied to deconvolution maps effectively suppresses
noise at small scales. The unsuppressed noise in the PR3 map is clearly visible in the difference map.

Another effect is the rescaling of the signal due to beam efficiency.
The signal collected by the far sidelobe component of the beam is removed as part of the calibration process,
and it is effectively lost.  
The lost power is accounted for in the beam model.
The beam efficiencies for LFI radiometer beams are given in \cite{planck2014-a05}.
The combined efficiency of main and near sidelobe beam components varies in the range
98.99--99.14\,\% for 30\,GHz, in 99.75--99.79\,\%  for 44\,GHz, and in 98.94--99.34\,\% for 70\,GHz.
The deconvolution process effectively scales the signal up by a corresponding factor,
compensating for the missing power.
As a consequence, we see a faint copy of the galaxy image in the difference map.

\begin{figure}
\includegraphics[trim={0.0cm 0 0 0}, width=8.8cm]{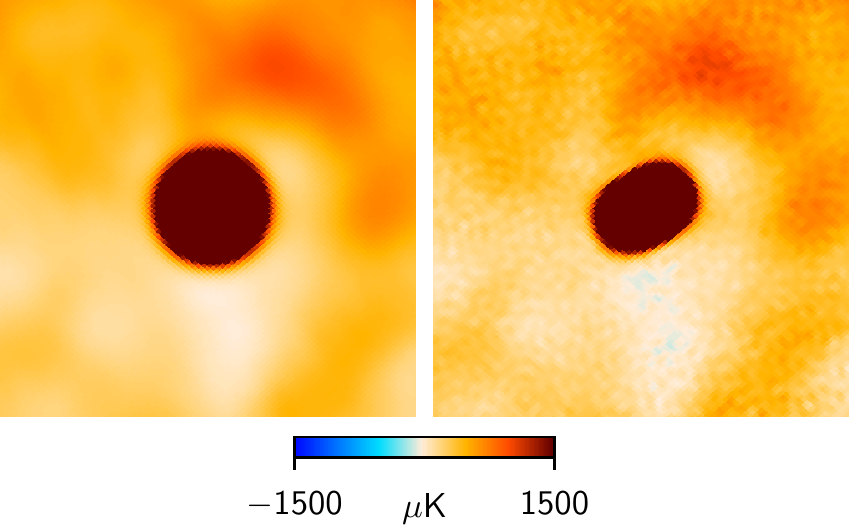}
\caption{A 10$^\circ$ patch around the Crab Nebula in the 30-GHz temperature map.
{\it Left}: deconvolved. {\it Right}: PR3 30-GHz LFI map.}
\label{fig:Crab_temperature}
\end{figure}

A zoom into a point source reveals the symmetrization of the effective beam.
As illustration, we show in Fig.~\ref{fig:Crab_temperature} the Crab Nebula,
as seen in the deconvolved 30-GHz temperature map
and in the corresponding undeconvolved map.
 In this figure, one can also see the effect deconvolution has on residual noise.
The weak striping visible in the upper left corner of the deconvolved map is noise,
which becomes correlated in the direction orthogonal to the direction of beam elongation.


\begin{figure}
\includegraphics[trim={0.0cm 0 0 0}, width=8.8cm]{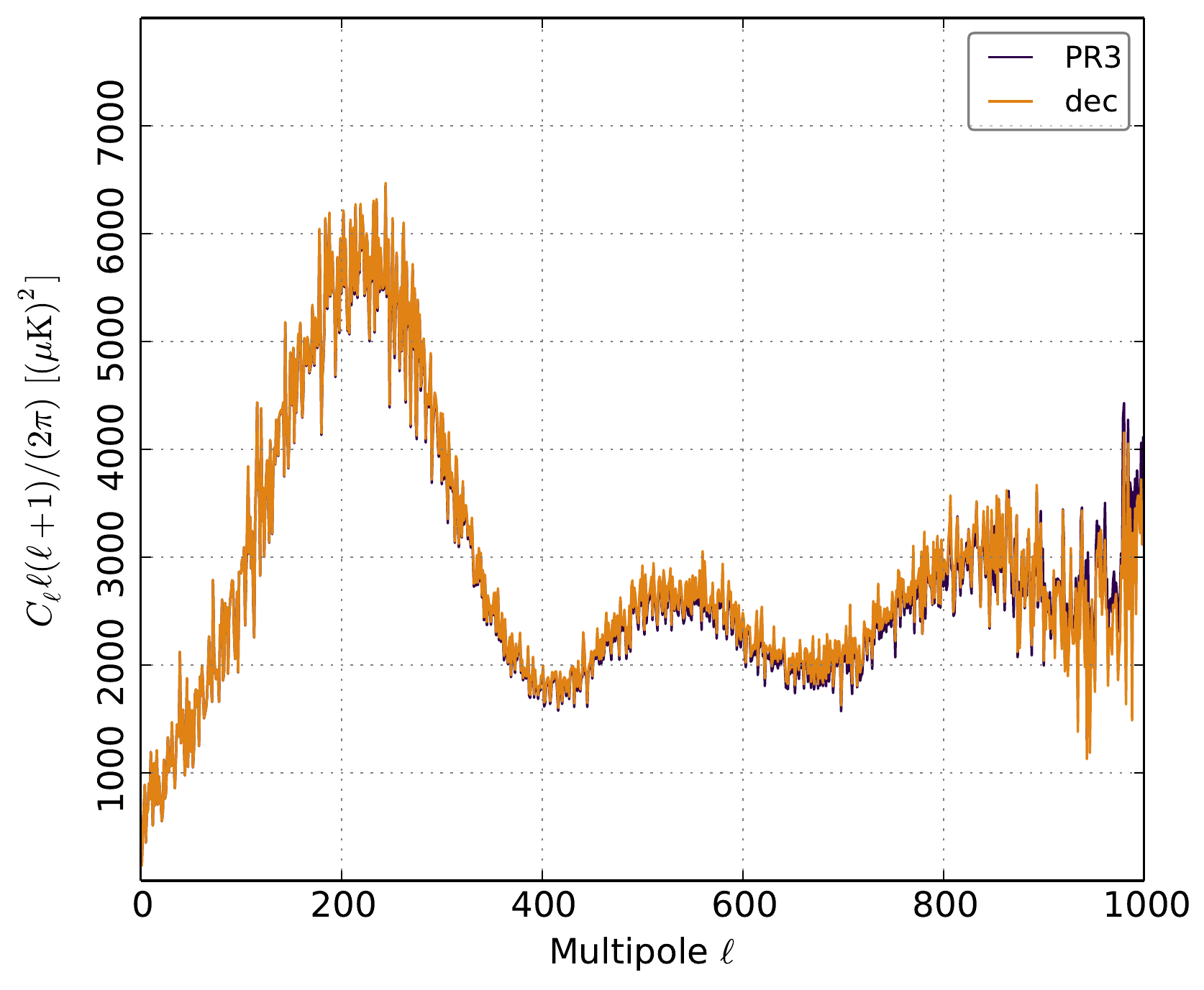}
\caption{{Half-ring cross-spectra in $TT$, from masked PR3 and deconvolution maps at 70 GHz,
after correction for for mask and beam window. }}
\label{fig:polspice}
\end{figure}

\subsection{Survey difference}
\label{sec:survey_difference}

{As a first validation test we produce temperature cross-spectra 
from PR3 and deconvolved half-ring maps at 70 GHz.  
We first smooth the PR3 map by a Gaussian FWHM=16' window to
match the final smoothing of the deconvolution map. 
We perform no component separation, but we apply a mask with 88.4\%
sky coverage to exclude regions of strong foreground emission.
We use {\sc PolSpice} \citep{chon2004}
to correct for the mode-coupling caused by the masking.
We divide the PR3 spectrum by the effective window function that is
 released with the instrument model, and by the 16' smoothing window.
The deconvolution spectrum is divided by the Gaussian 20' smoothing window,
and by the pixel window.
In the case of the PR3 map we do not correct for the pixel window,
since it is already included in the effective beam window.
The test shows a good agreement between the spectra,
as shown in Fig. \ref{fig:polspice}
}

As we see above, the difference between deconvolved and PR3 full mission maps
{arises largely from} effects not directly related to beam asymmetry.
To see the effects of actual beam symmetrization more clearly, we now proceed to look at survey difference maps. 
A survey in this context refers to a period of six months, during which the scan axis of \Planck\ 
rotates by $180^\circ$, and detectors scan almost the complete sky.  
During the subsequent 6 months the sky is scanned again, this time with effectively inverted beam orientation.
When we take the difference between odd and even single survey maps, most of the sky signal cancels out,
apart from a residual that arises from beam shape mismatch.

Beam deconvolution is expected to remove the beam residuals,
leaving only noise in the survey difference maps,
since the deconvolved maps have identical effective beams.
With beam residuals out of the way,
remaining signal residuals provide valuable information on other systematic
artefacts in the data.

Single surveys do not provide complete sky coverage.  The coverage varies between 79\,\% and 98\,\%, 
depending on survey and frequency \citep{planck2014-a07}.
The \artdeco\ deconvolver operates with harmonic coefficients, which necessarily represent the entire sky.
If data are not available for the full sky, the correct sky signal is well recovered in the region 
covered by the measurements. In the missing region (and close to its boundary),  
we obtain a solution that is based on extrapolation of the recovered signal.

We construct an odd-even difference map from the first four surveys as combination S1+S3-S2-S4.
We leave out the last four surveys, which have lower sky coverage \citep{planck2016-l02}.
To suppress noise and to bring out the beam 
residuals more clearly, we build the maps at $1^\circ$ resolution (FWHM).  
We apply smoothing with a Gaussian beam, the width of which is chosen to make
 the combined width of the smoothing kernel 
and the average detector beam equal to $1^\circ$.   
The required additional smoothing is 50\parcm63, 53\parcm58, and 58\parcm53 for 30, 44, and 70\,GHz, respectively.


\begin{figure*}
\begin{center}
\includegraphics[width=18cm]{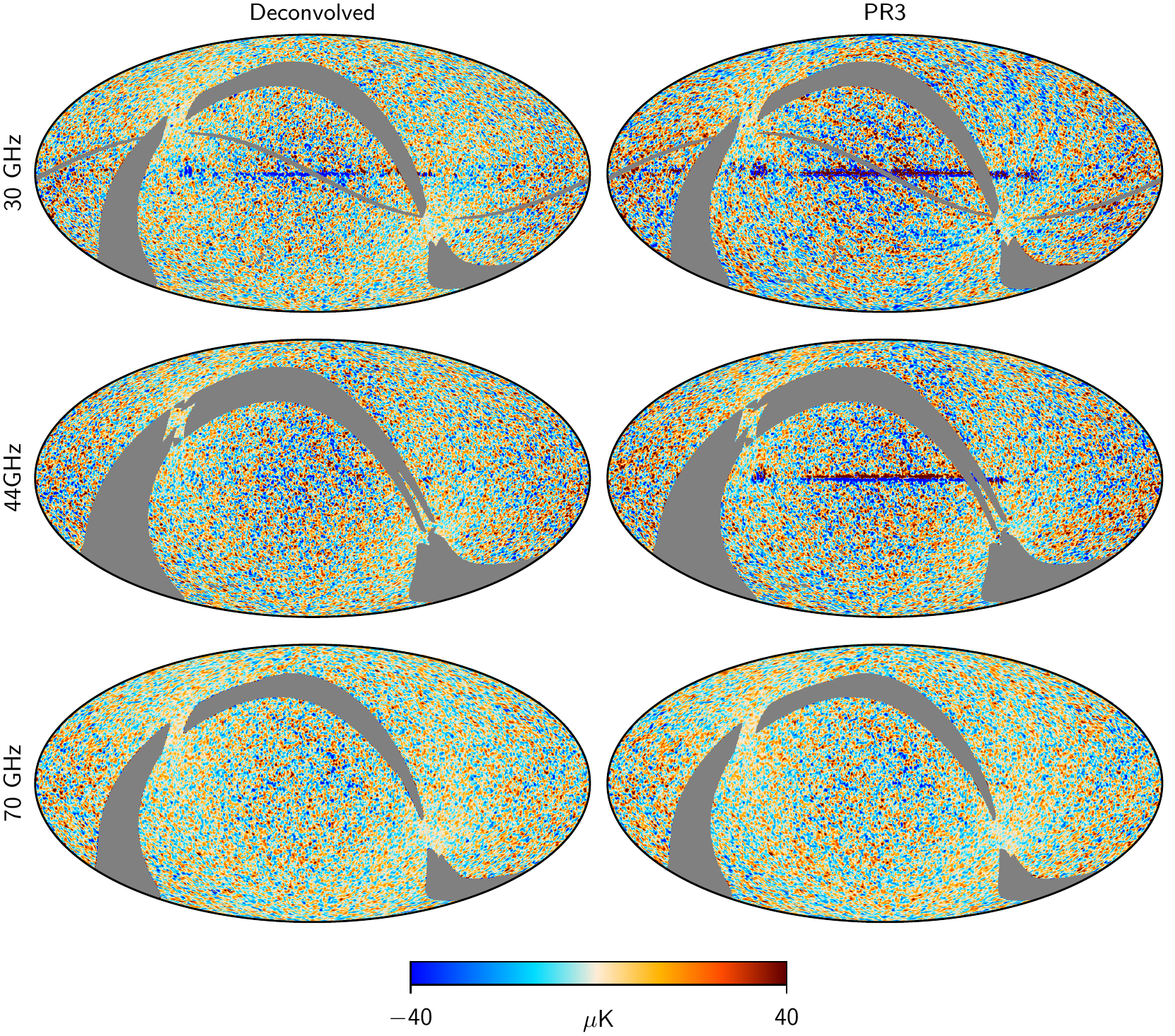}
\caption{Survey difference $T$ maps S1+S3-S2-S4.
The maps are smoothed to $1^\circ$ resolution.
Deconvolution reduces beam shape residuals, revealing other signal residuals.
}
\label{fig:odd-even}
\end{center}
\end{figure*}

The masked odd-even survey difference $T$ maps are shown in Fig.~\ref{fig:odd-even}.
The pixels outside the mask are covered in all four surveys.
For each frequency, we show the deconvolved version on the left, and the corresponding PR3 map on the right.
We note that the PR3 maps are destriped with the independent destriping option, as discussed in Sect. \ref{sec:inputTOI},
while the deconvolved maps are destriped with the full destriping option.
{This follows from the fact that the destriped ring objects, which are the input to deconvolution,
are only available from full-mission runs.}
The deconvolved maps thus contain less correlated noise to begin with.

In the Galactic region, the PR3 30-GHz maps show signal residuals with typical beam residual characteristics, 
the signal varying between positive and negative values. Point sources give rise to `butterfly' shapes. 
Deconvolution reduces the residuals significantly, but does not remove them completely. It is likely that there are other systematic 
effects than beam at play here. As described in  Planck Collaboration II (2018), a possible cause is the incomplete
convergence in the calibration iterative approach.

The effect of deconvolution is clearest for 44\,GHz:  PR3 maps show strong beam residuals,
arising from beam shape mismatch between the two survey combinations.
Deconvolution symmetrizes the effective beam, and removes the residuals almost perfectly.
At 70\,GHz, the residuals are small to start with and no signal residuals are visible in either map.


\subsection{Polarization maps}

Beam-shape mismatch is a source of leakage of temperature signal into polarization.
When the same point on the sky is observed twice by the same detector, but in different beam orientations,
the two measurements collect slightly different temperature signals.
The usual map binning operation interprets the difference as a polarization signal.
Because the temperature signal is much stronger than the polarization signal,
the leakage can affect the polarization measurements significantly.
According to \cite{planck2014-a05}, beam leakage can contribute as much as 15\,\%
of the polarization signal at 70\,GHz.

Simulations show that beam deconvolution can drastically reduce temperature leakage
 \citep{keihanen2012}.
In real \Planck\ measurements, however, beam mismatch is not the only effect.
Another, and often more significant, source of temperature leakage is bandpass mismatch,
coming from the fact that
different radiometers have slightly different frequency responses.
Two radiometers record different foreground temperature signals,
and the difference is interpreted incorrectly as a polarization signal.
This affects deconvolved and un-deconvolved maps alike,
and partially hides the beam correction.

\begin{figure*}
\includegraphics[width=18.0cm]{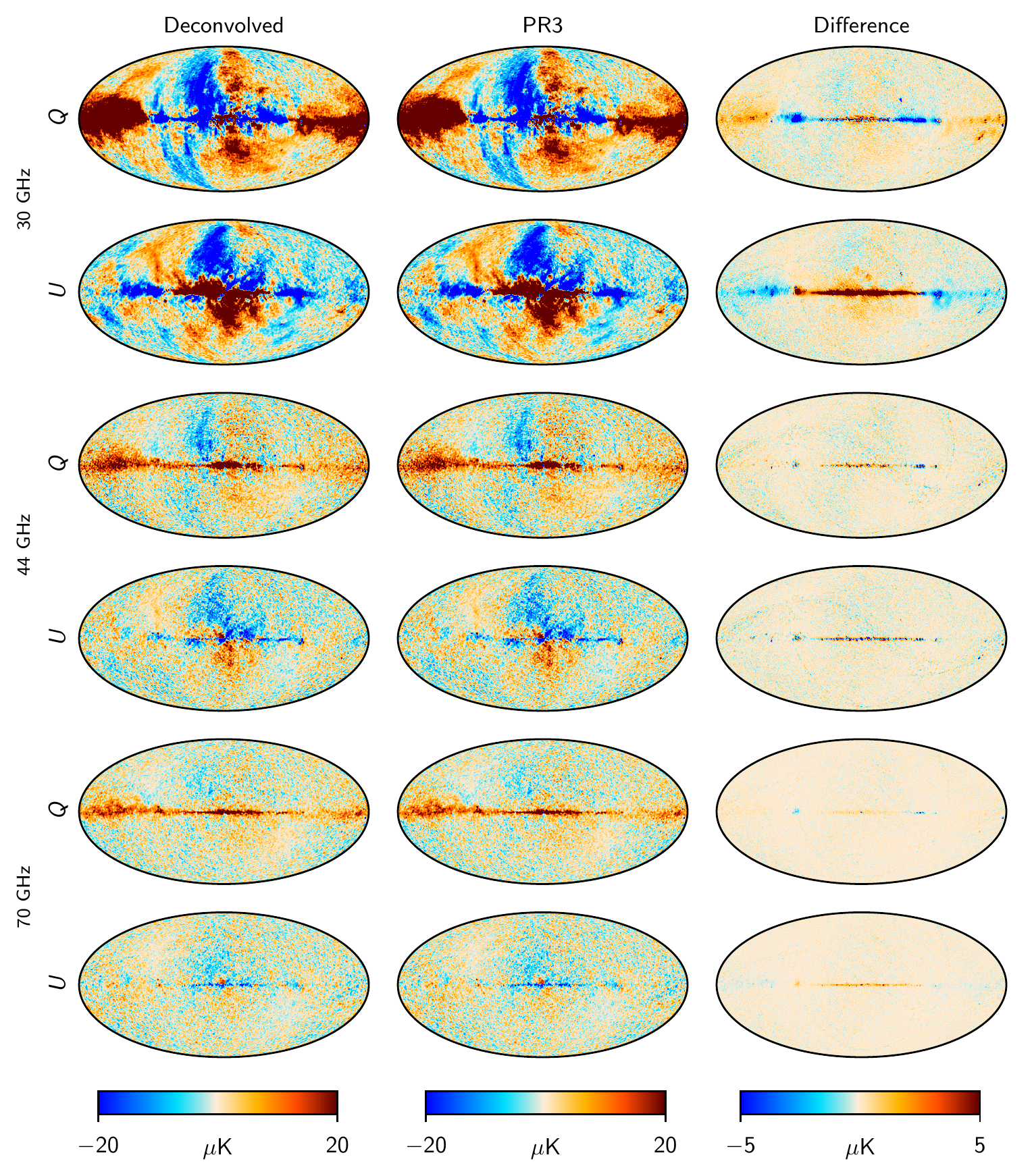}
\caption{Deconvolved (left) and PR3 (middle) polarization maps for LFI channels,
and their differences (right).
From top down: 30\,GHz; 44\,GHz; and 70\,GHz.
These maps are smoothed to $1^\circ$ resolution (FWHM) to suppress noise.
The maps are not corrected for bandpass leakage.
}
\label{fig:polarization_maps}
\end{figure*}

Figure \ref{fig:polarization_maps} shows the $Q$ and $U$ polarization component maps for LFI frequencies.
We compare deconvolved and PR3 LFI maps, and plot their difference in the third column.
Since we are interested in the effect of beam deconvolution,
we use in this comparison PR3 maps that are not corrected for bandpass leakage.
The maps are smoothed to $1^\circ$ resolution to suppress the noise.
The smoothing procedure is the same as the one applied to single survey maps in Sect. \ref{sec:survey_difference}.
The difference map is plotted over a 2 $\mu$K scale to bring out the structure above the Galactic plane.

We expect the difference maps to include the temperature leakage through beam shape mismatch,
but this is difficult to verify, since the difference maps are dominated by other effects.
As in the case with temperature maps, deconvolution scales the map up to compensate for the missing power 
absorbed by the far sidelobe beam component.
Since {the missing power} is slightly different for different radiometers,
this also changes the relative weighting of radiometers, and hence the bandpass leakage pattern.
As a result, the difference map between deconvolved
and PR3 LFI polarization maps
is a combination of beam leakage and changes in scaling and in the bandpass leakage pattern.

Although the temperature-to-polarization leakage is dominated by bandpass mismatch,
it still makes sense to remove the component that comes from beam asymmetry.
Bandpass leakage is relatively well understood and can be corrected for
more easily than beam leakage.
The bandpass correction for the 2018 release is described in \cite{planck2016-l02}.
Furthermore, bandpass leakage only affects foreground emission,
while beam leakage distorts both foregrounds and the CMB.

{We do not show survey difference maps in polarization,
since the beam effects are obscured by effects of bandpass mismatch.
The hit count distribution per radiometer varies between surveys,
and so does the bandpass leakage pattern.
}


\subsection{IQUSS deconvolution}

As an alternative to the usual deconvolution procedure, we have applied a procedure
similar to the `{\it IQUSS}' mapmaking presented in \cite{planck2016-l02}.
We split the harmonic temperature coefficients $a_{T\ell m}$ further
into a component that represents the average temperature signal,
and components that represent the signal difference due to different frequency responses
between a radiometer pair.
For radiometers M and S of horn X,
\begin{eqnarray}
a^T_{\ell m,XM} &=&  a^T_{\ell m} +S^T_{\ell m,X}  \nonumber \\
a^T_{\ell m,XS} &=&  a^T_{\ell m} -S^T_{\ell m,X} \,,
\end{eqnarray}
where $a^T_{\ell m}$ represents the average sky signal, and $S^T_{\ell m,X}$
is a spurious signal map for horn X.  The opposite signs are chosen so that 
the spurious signal will absorb the
signal difference from bandpass mismatch, preventing it from leaking into polarization.
We note that we do not include spurious 
components between horns because signal differences between horns do not
contribute to polarization.
We now solve for the spurious components along with the
usual  $a^T_{\ell m}$, $a^E_{\ell m}$, and $a^B_{\ell m}$.

The spurious components obey the rotation properties of a temperature map.
They can be distinguished from the polarization components
if the sky is scanned in sufficiently different beam orientations during the mission.
Unfortunately, for Planck this is true only for part of the sky
due to the scanning strategy, which was optimised for temperature, not polarization. 
Consequently, the polarization maps we obtain from spurious fitting
are very noisy compared to the usual deconvolution with three sky components.
The same applies naturally to the PR3 {\it IQUSS} maps \citep{planck2016-l02}.

Deconvolved {\it IQUSS} maps constructed from PR2 data were used for the analysis of Tau A polarization
in the 2015 data release, as described in \cite{planck2014-a35}.
Due to the experimental nature of this method this approach  was not extensively 
used in other Planck analyses.

The {\it IQUSS} maps could be used to construct a deconvolution-specific bandpass correction
following the same procedure as the bandpass correction for undeconvolved maps;
however, more analysis would be needed to implement this.  
At present, the best way of correcting the bandpass leakage in deconvolved maps is to apply
the un-deconvolved correction maps directly.

\subsection{Point sources}

The 2015 \Planck\ release included an extensive catalogue of compact sources \citep{planck2014-a35}.
The catalogue is built on the PR3 (undeconvolved) frequency maps.
Deconvolved maps provide valuable complementary information on the properties of sources.

In the following, we demonstrate the fidelity of the deconvolution process 
by showing that it efficiently reduces the asymmetry of a point source image.
We take the ellipticity of the image as figure of merit.
We pick the quasar 3C279 for demonstration.
We consider also TauA (the Crab Nebula), which is an important calibration source in CMB experiments.
TauA can be assumed to be point-like at the lowest LFI channels, but becomes extended at HFI channels \citep{planck2014-a35,ritacco2018}

We determine the ellipticity of both sources in the pixel domain
by fitting a two-dimensional Gaussian shape to the source image.
To reduce the contamination from background signals,
we fit a constant offset along with the shape.
We determine the ellipticity as the ratio of the two major axes.
The ellipticity is therefore always above 1,
the value corresponding to a circular source.
Table~\ref{tab:sources} gives the measured ellipticies for the selected sources, 
for deconvolved and PR3 maps.
For each source we list the ellipticity of the source,
as detemined from the undeconvolved and deconvolved maps.
As expected, deconvolution strongly reduces the ellipticity of the source image.

\begin{table}
\caption{Properties of selected point sources.  The ellipticity is determined from the undeconvolved PR3 map and from the deconvolved map.}
\label{tab:sources}
\vskip -7mm
\footnotesize
\setbox\tablebox=\vbox{
\newdimen\digitwidth
\setbox0=\hbox{\rm 0}
\digitwidth=\wd0
\catcode`*=\active
\def*{\kern\digitwidth}
\newdimen\signwidth
\setbox0=\hbox{+}
\signwidth=\wd0
\catcode`!=\active
\def!{\kern\signwidth}
\newdimen\decimalwidth
\setbox0=\hbox{.}
\decimalwidth=\wd0
\catcode`@=\active
\def@{\kern\signwidth}
\halign{\hbox to 0.7in{#\leaderfil}\tabskip=1.5em&
    \hfil#\hfil\tabskip=1em&
    \hfil#\hfil\tabskip=1.4em&
    \hfil#\hfil&
    \hfil#\hfil\tabskip=1em&
    \hfil#\hfil\tabskip=1.4em&
    \hfil#\hfil\tabskip=0em\cr
\noalign{\doubleline}
\omit&\multispan2\hfil Gal Coord\hfil&$\nu$&\multispan2\hfil Ellipticity\hfil\cr
\noalign{\vskip -2pt}
\omit&\multispan2\hrulefill&&\multispan2\hrulefill\cr
\noalign{\vskip 2pt}
\omit\hfil Source\hfil&\omit\hfil $l$\hfil&\omit\hfil $b$\hfil&[GHz]&Undec&Dec\cr 
\noalign{\vskip 5pt\hrule\vskip 5pt}
 \noalign{\vskip 4pt}
TauA& 184\deg5& *$-$5\pdeg8&      30&  1.3274& 1.0042&   \cr
 \omit                          &       &    &      44&  1.0067& 1.0026&  \cr
 \omit                          &       &    &      70&  1.2217& 1.0096&  \cr
 \noalign{\vskip 4pt}
3C279& 305\pdeg1&       !57\pdeg1&  30&  1.3247& 1.0038& \cr
 \omit                              &       &    &   44&  1.0135& 1.0123&  \cr
\omit                               &       &    &   70&  1.2365& 1.0130&  \cr
\noalign{\vskip 4pt\hrule\vskip 2pt}
}}
\endPlancktable
\end{table}


\section{Low-resolution analysis}
\label{sec:lowresolution}

Computation of the CMB angular power spectrum and subsequent cosmological analysis
is complicated by noise,  and a detailed description of the residual noise in the maps
is therefore required. This is usually given in the form of a pixel-pixel noise covariance matrix (NCVM).
For reviews of power spectrum methods designed for \Planck, see \cite{gruppuso2009} and \cite{planck2014-a13}.

A properly constructed noise covariance matrix gives a full description of noise correlation
between any pair of low-resolution pixels. 
The 2018 \Planck\ data release includes low-resolution maps and corresponding 
noise covariance matrices at \healpix\ resolution \nside=16,
This corresponds to  3\deg7 spatial resolution.
The low-resolution maps are constructed from their high-resolution counterparts through 
a smoothing and downgrading process,
as described in \cite{planck2014-a07}.
Each low-resolution map consists of 3072 sky pixels.
Each pixel further contains estimates for the three Stokes parameters $I$, $Q$, and $U$.
The corresponding NCVM objects contain $9\times3072\times3072\approx8.5\times10^7$ numbers.
Increasing the resolution to $\nside=32$ would increase the matrix size further by a factor of 16,
which becomes computationally prohibitive.

\subsection{Low-resolution deconvolution}

Deconvolution offers an alternative way of extracting the low-multipole information from CMB data.
The division into low- and high-multipole regimes occurs in a natural way here,
since deconvolution operates in harmonic space.  

We construct deconvolved low-resolution maps from LFI data as follows. 
For harmonic analysis, we simply take the harmonic coefficients 
we have from full-resolution deconvolution,
and pick the lowest multipoles for the low-resolution data set.
To construct low-resolution pixel maps, we apply
a $\hbox{FWHM} = 40'$ Gaussian beam to the coefficients, 
and construct maps at resolution $\nside=16$.
The smoothing width and resolution were chosen so that they correspond to the PR3
procedure as closely as possible.

We construct noise covariance matrices (NCVM)
that describe the noise in the deconvolved low-resolution maps,
both in harmonic and pixel space.
The construction of the NCVM is described in detail in \cite{keihanen2015}.
Required inputs are pointing,  a model for the power spectral density (PSD) 
of noise in the time domain, and beams.
The NCVM matrix essentially accounts for two aspects:  
the correlated residual noise that remains in the data after destriping;
and the effect of the deconvolution process on it.
Deconvolution modifies the noise structure,
but the effect is seen mostly at high multipoles.
Our harmonic NCVM describes the noise correlation between
any two $a_{slm}$ elements up to $\ell=50$.
The pixel-space NCVM describes the correlation between two $\nside = 16$ pixels.
The noise model we use adopts the noise estimates of the
official DPC ({Data Processing Center}) pipeline.  These noise estimates are provided in the form of a PSD, 
as a function of frequency, up to half the sampling frequency of each channel.
These estimates include one spectrum per day for each detector.
The PSDs are estimated from the data, and vary strongly with frequency.

For 30 and 70\,GHz, we use the raw PSDs as they are.
To construct the NCVM, we need to divide each PSD into a white noise component 
and a correlated component.
To do this, we arrange the spectrum as a function of inverse frequency,
and perform a linear fit to the last ten points corresponding to the highest frequencies.
We extrapolate the fit to $1/f=0$, which we take as the white noise power.
We subtract this from the full spectrum to obtain the correlated component.
The procedure is exact for a pure power-law spectrum with slope equal to $-1$.

The procedure described above works well for 30\,GHz and 70\,GHz,
but at 44\,GHz it yields unphysical negative values for the correlated power
due to the large variations in the spectra.
 At 44\,GHz, we therefore use a modified procedure,
where a power-law spectrum is first fitted to the raw PSD
to yield a smooth spectrum before applying the same extrapolation procedure
as for 30\,GHz and 70\,GHz.


\subsection{NCVM validation in harmonic space}

The PR3 data release includes half-ring maps.
These are constructed by splitting each pointing period in two, constructing two data sets from these,
and applying the usual flagging and map-making procedure to both sets.
The result is two half-ring (HR) maps with nearly identical signal content,
but independent noise.
The half-ring half-difference (HRHD) map can be constructed as the difference between
the two HR maps,  weighted by the actual hit count, as explained in  \cite{planck2013-p02}.
In an ideal situation this consists of pure noise,
with structure similar to that of the noise contained in the full map.
However, small differences arise because the noise correlations on the longest timescales
are not tracked by the half-ring process,
and the spatial sample distributions in the two halves are not identical.
Despite its limitations, the HRHD maps provide the best estimate of the actual noise level in the data;
these are independent of any noise modelling and are based entirely on the data.

The half-ring data sets for Planck LFI are available in the form of ring objects.
This allows us to run the deconvolution process on the two HRs,
and to form a HRHD harmonic vector, following a logic similar to that behind the HRHD maps.
A harmonic HRHD vector is thus a set of $a_{slm}$ coefficients
consisting of pure noise with properties similar to the noise in the actual deconvolved $a_{slm}$.  
We validate the harmonic NCVM matrix by comparing its predictions to this HRHD noise estimate.

We readily obtain the {\it noise bias} from the diagonal of the harmonic NCVM.
The noise bias is a prediction for the angular power of residual noise.
As a first test we compare the noise bias with the angular power spectrum of HRHD noise,
constructed directly from the deconvolved harmonic coefficients as

\begin{equation}
{\cal C}_{XX}(\ell) = \frac{1}{2\ell+1}\sum_m a_{Xlm}^\ast a_{Xlm}\,,
\end{equation}
where $X$ stands for $T,E$, or $B$.

\begin{figure*}
\includegraphics[width=18cm]{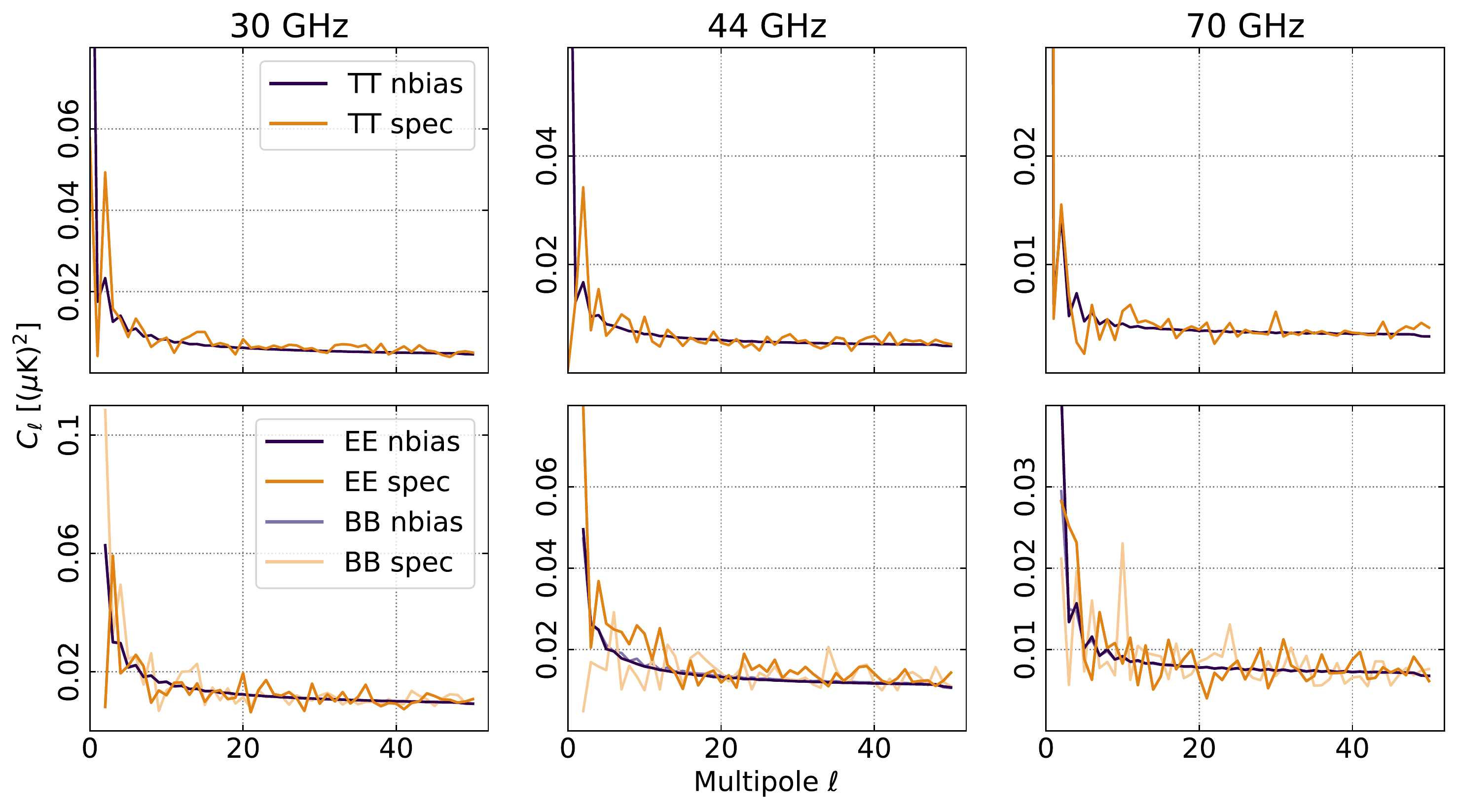}
\caption{Noise bias from NCVM, and the HRHD noise estimate, for 30, 44, and 70\,GHz data,
{\it Top}: $TT$ spectrum.  {\it Bottom}: $EE$ and $BB$ spectra.}
\label{fig:noisebias}
\end{figure*}

Figure \ref{fig:noisebias} shows the spectra and the corresponding noise bias for 
LFI frequencies for $TT$, $EE$, and $BB$.
In general, the noise bias agrees well with the HRHD spectrum,
but there are also differences.
In particular, at 44\,GHz there is an asymmetry between $EE$ and $BB$ spectra,
the $EE$ spectrum being above the noise bias, the $BB$ spectrum falling below it.
The same asymmetry is already present in the un-deconvolved data,
as can be seen in figure 18 of \cite{planck2016-l02},
as well as in the 2015 data \citep{planck2014-a03}.
The 44\,GHz channel is different from 30\,GHz and 70\,GHz as it involves the lone horn 24, which has no complementary pair with orthogonal
polarization sensitivity, and which has different beam properties
from the other 44-GHz horns. All these aspects are nevertheless accounted for in the NCVM
and in the noise bias derived from it. The noise bias shows only a very weak asymmetry
between $EE$ and $BB$, as we see in the middle panel of Fig.~\ref{fig:noisebias}.
Monte Carlo simulations in \cite{planck2016-l02} also fail to reproduce the asymmetry
that in any case is well inside 1 sigma.

\begin{figure}
\includegraphics[width=8.8cm]{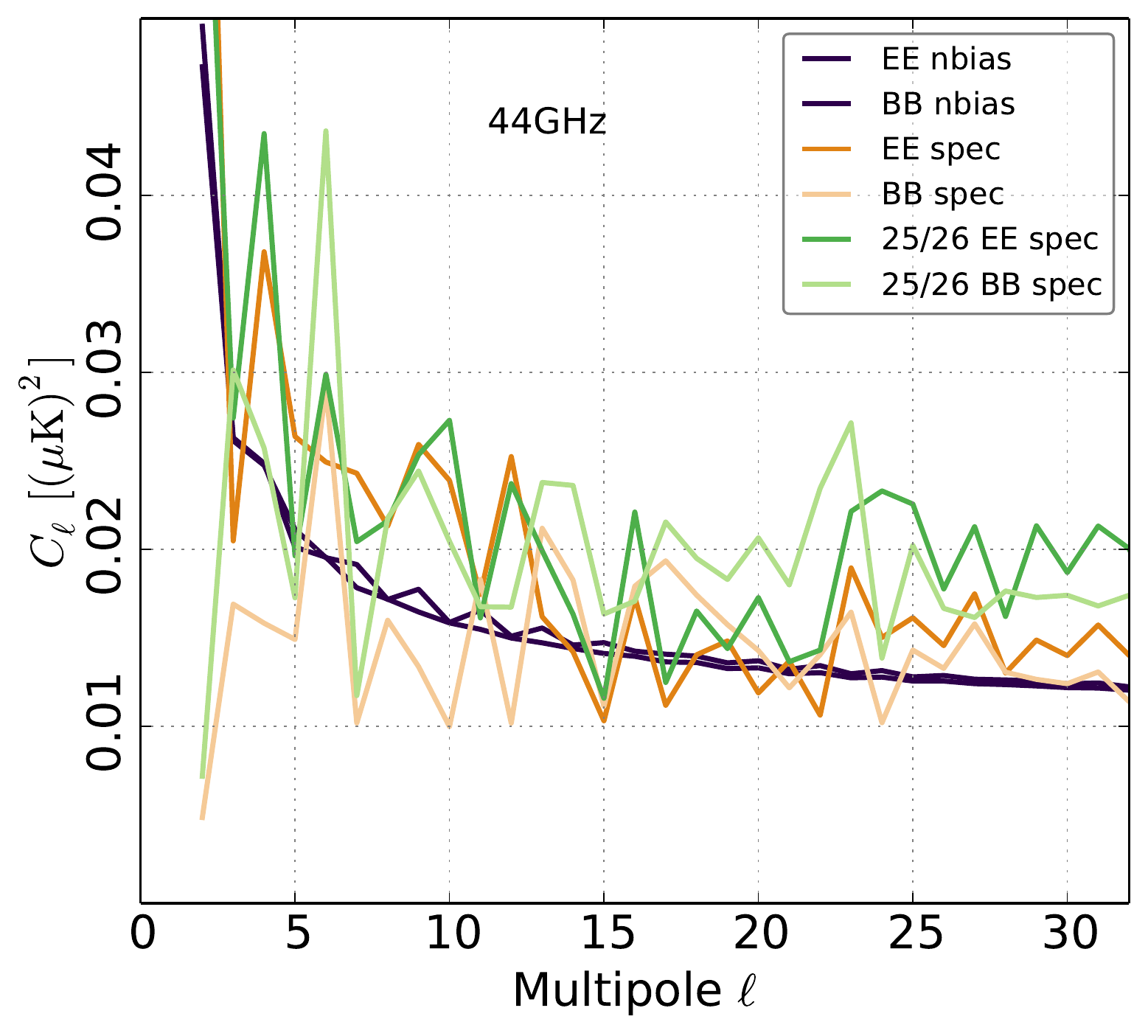}
\caption{Noise bias from NCVM  for full 44\,GHz data,
and HRHD noise estimate for full 44\,GHz data and for 25/26 horn combination.
}
\label{fig:noisebias44}
\end{figure}

{We replot the 44\,GHz polarization spectra in Fig. \ref{fig:noisebias44},
together with the HRHD spectra derived from the LFI 25/26 horn combination,
excluding the individual horn 24.  With horn 24 excluded, the noise level is higher, 
roughly by a factor of 3/2 as expected.
The asymmetry between EE and BB is reduced, suggesting that the asymmetry is related to 
the unusual detector combination -- the horn pair plus individual horn -- of the 44 GHz channel.
The exact mechanism however remains unclear.
}

It is interesting to compare the deconvolved HRHD spectra with their un-deconvolved counterparts.
The structure of peaks and troughs is very similar in both. For instance,
the high peak at $\ell$=10 in the 70-GHz $BB$ spectrum is also present
in figure 18 of \cite{planck2016-l02}.


A $\chi^2$ test provides a more stringent validation test for the NCVM.
We specifically compute the measure

\begin{equation}
\chi^2 = \frac{\vec{a}^{\sf T} N^{-1} \vec{a}}{N_{\rm dof}}   \label{chi2general}\,,
\end{equation}
where $N$ is the covariance matrix, $\vec a$ is the corresponding HRHD object,
and $N_{\rm dof}$ is the number of degrees of freedom. 
Dividing by $N_{\rm dof}$ yields the {\em reduced} $\chi^2$ test.
If the matrix correctly models the noise properties, we have $\langle \vec a \vec a^{\sf T}\rangle=N$,
and the reduced $\chi^2$ test is expected to return a value close to 1.
The value is not limited from below however.  Therefore, a simple overestimation of the overall noise level
is able to compensate for another error, and this must be kept in mind when interpreting the results.
Despite its limitations, the $\chi^2$ test provides a useful measure of NCVM quality.

The $\chi^2$ results for the harmonic NCVM are given in Table~\ref{table:chialm}.
We compute the $\chi^2$ value between the full mission NCVM of each frequency channel
and the corresponding HRHD noise object.  We perform the test separately for the $TT$, $EE$, and $BB$ blocks.
The number of degrees of freedom is $N_{\rm dof}=(\ell_{\rm max}+1)^2$,
which for $\ell_{\rm max}=50$ yields $N_{\rm dof}=2601$.
We repeat the $\chi^2$ test for a submatrix that covers multipoles from zero to $\ell=40$ or 30.

\begin{table}
\caption{Reduced $\chi^2$ results in harmonic space.  We test the $TT$, $EE$, and $BB$ blocks of the harmonic NCVM against the corresponding half-ring noise estimate.  The calculation includes all detectors and all available data for the frequency channel in question.  The expected one-sigma deviations from unity are 0.0196, 0.0244, and 0.0323 for \lmax$=50$, 40, and 30, respectively.}
\label{table:chialm}
\footnotesize
\vskip -4mm
\setbox\tablebox=\vbox{
\newdimen\digitwidth
\setbox0=\hbox{\rm 0}
\digitwidth=\wd0
\catcode`*=\active
\def*{\kern\digitwidth}
\newdimen\signwidth
\setbox0=\hbox{+}
\signwidth=\wd0
\catcode`!=\active
\def!{\kern\signwidth}
\newdimen\decimalwidth
\setbox0=\hbox{.}
\decimalwidth=\wd0
\catcode`@=\active
\def@{\kern\signwidth}
\halign{\hbox to 0.9in{#\leaderfil}\tabskip=2.0em&
    \hfil#\hfil\tabskip=2em&
    \hfil#\hfil\tabskip=2em&
    \hfil#\hfil\tabskip=0em\cr
\noalign{\doubleline}
\omit&\multispan3\hfil Reduced $\chi^2$\hfil\cr
\noalign{\vskip -2pt}
\omit&\multispan3\hrulefill\cr
\noalign{\vskip 2pt}
\omit\hfil Channel, $\ell$\hfil&$TT$&$EE$&$BB$\cr 
\noalign{\vskip 5pt\hrule\vskip 5pt}
 \noalign{\vskip 4pt}
\omit \bf 30\,GHz\hfil\cr
\hglue 1em\lmax$=50$& 1.1132& 1.0458& 1.0498\cr
\hglue 1em\lmax$=40$& 1.1134& 1.0268& 1.0101\cr
\hglue 1em\lmax$=30$& 1.0739& 1.0296& 1.0117\cr
\noalign{\vskip 4pt}
\omit \bf 44\,GHz\hfil\cr
\hglue 1em\lmax$=50$& 1.0752& 1.1611& 1.0829\cr
\hglue 1em\lmax$=40$& 1.0422& 1.1651& 1.0752\cr
\hglue 1em\lmax$=30$& 1.0406& 1.1442& 1.0259\cr
\noalign{\vskip 4pt}
\omit \bf 70\,GHz\hfil\cr
\hglue 1em\lmax$=50$& 1.1029& 1.0451& 1.0412\cr
\hglue 1em\lmax$=40$&  1.0523& 1.0330& 1.0442\cr
\hglue 1em\lmax$=30$&  1.0877& 1.0050& 1.0777\cr
\noalign{\vskip 4pt\hrule\vskip 2pt}
}}
\endPlancktable
\end{table}

Consider first the 30\,GHz and 70\,GHz data.
The NCVM appears to model the residuals rather well in polarization.
The deviations from unity for $EE$ and $BB$ are within $2.2\sigma$ statistical variation
($\sigma=0.0196$ for $\ell=50$).
However, all the results are above unity, indicating that the NCVM may be slightly underestimating 
the residual noise.

At 30\,GHz  the results improve further  when the highest multipoles are excluded.
However, $TT$ results are much poorer, especially at 30\,GHz.
A misestimation should affect temperature and polarization in roughly the same way, and therefore the weaker results in temperature indicate that the HRHD maps contain signal residuals.
The likely source is signal variations within a sky pixel in the Galactic region where signal gradients are strong.
Signal gradients distort the destriping solution, which assumes that the signal is
constant within a pixel. The effect depends on the exact scanning pattern,
which is different for the two HRs. 
However, a pair of \Planck\ detectors sharing a horn follow very closely the same path on the sky.
Since the polarization signal is determined from the signal differences of such pairs,
the signal gradients tend to cancel in the polarization component of the HRHD map.
The effect is not restricted to the pixels with strong gradients, but is redistributed over the sky through the destriping process.
The error can be reduced with the help of a properly chosen destriping mask.
The choice of the mask however is a trade-off because too wide a mask will remove too much of the data.

At 44\,GHz we observe the same suspicious excess $EE$ power that we already observed in the HRHD spectrum.
The excess is insensitive to the choice of \lmax.
In contrast to 30 and 70\,GHz, the $\chi^2$ results are better in temperature than polarization.

\subsection{Validation in pixel space}

We now proceed with the validation in pixel space.
We produce low-resolution \healpix\ maps as follows. 
We take as a starting point the harmonic coefficients 
that we have as deconvolution output,
and smooth them with a Gaussian of {a 440\arcm (FWHM)}. 
The same smoothing width is adopted for the PR3 pipeline.
We then construct temperature and polarization maps at resolution $\nside=16$
using the standard \healpix\ tools.
The same operations are applied to each row and column of the harmonic NCVM
to obtain a pixel-pixel noise covariance matrix.

Moving to pixel space changes the number of degrees of freedom.
The harmonic NCVM with $\lmax=50$ has 2601 eigenmodes,
while a $\nside=16$ pixel-pixel NCVM has rank 3072.
Moving from harmonic to pixel space thus introduces 471 zero eigenmodes, 
resulting in a singular NCVM matrix.
There is nothing unphysical about this. 
It just tells us that these eigenmodes are not present in the data.
However, a $\chi^2$ test requires
that the matrix be inverted. We thus need to regularize the matrix in some way.

\begin{figure}
\includegraphics[width=8.8cm]{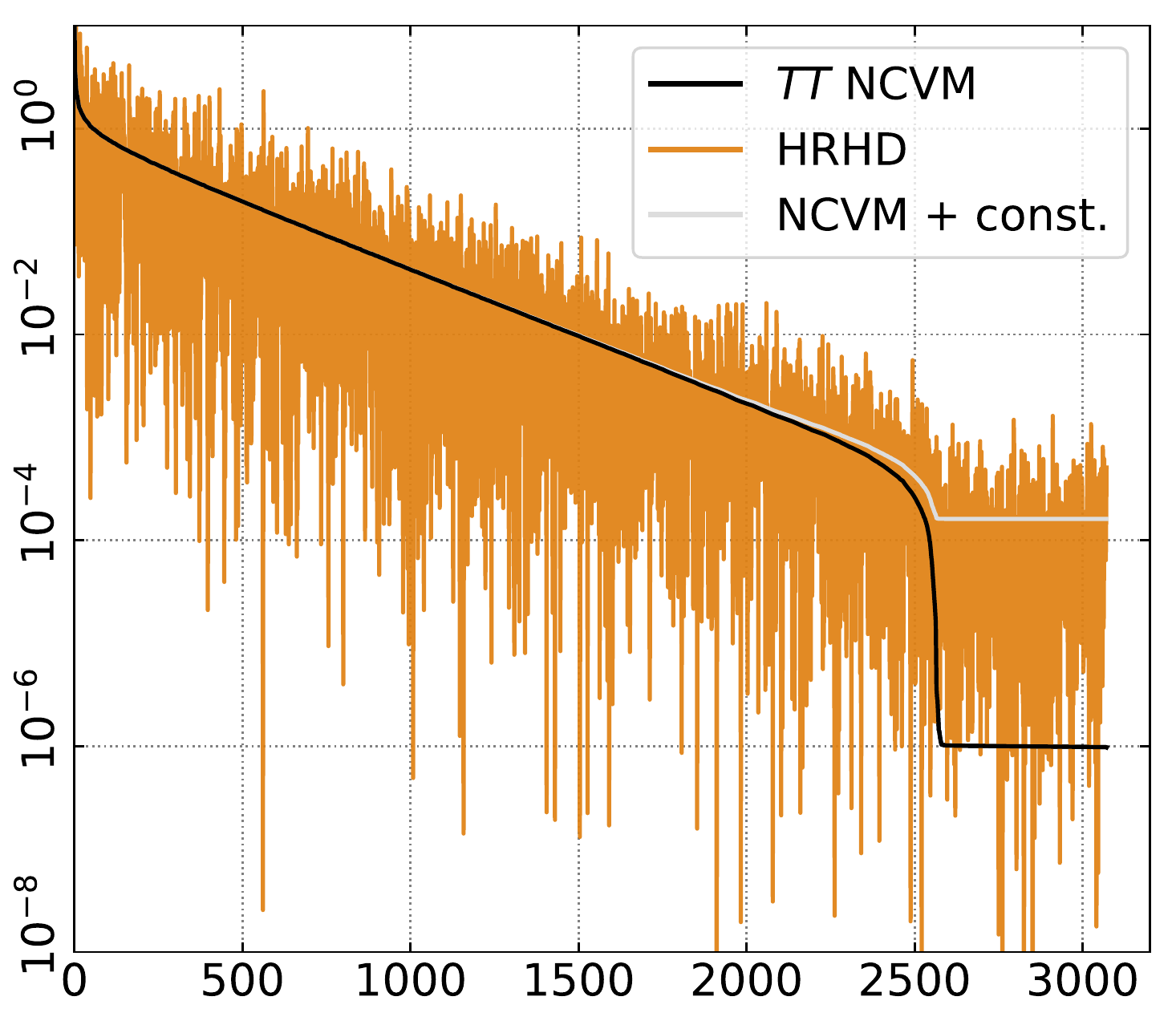}
\caption{Eigenvalue spectrum of the 30-GHz $TT$ NCVM (grey),
and corresponding decomposition of the HRHD map (orange).
We add a small constant of $10^{-6}\mu\mathrm{K}^2$ to the NCVM for numerical stability.
A larger phenomenological constant is added to the NCVM (light blue)
to account for the high-end plateau.
}
\label{fig:eigenvalues}
\end{figure}

To do this we start by constructing the eigenvalue decomposition of the pixel NCVM,
and rewrite matrix $N$ as

\begin{equation}
 N = \sum_j \lambda^2 \vec e_j \vec e_j^{\sf T} \ ,
\end{equation}
where $\vec e_j$ are the eigenvectors and $\lambda_j$ the corresponding eigenvalues.
Equation (\ref{chi2general}) takes the form

\begin{equation}
  \chi^2 = \frac{1}{N_{\rm dof}} \sum_j \frac{ (\vec e_j^{\sf T} \vec m)^2} {\lambda_j^2} \,,
\end{equation}
where $\vec m$ is the HRHD noise map under consideration.
In Fig.~\ref{fig:eigenvalues} we plot the eigenvalue spectrum $\lambda_j^2$ of the 30-GHz $TT$ matrix,
and the decomposition of the corresponding HRHD map, $ (\vec e_j^{\sf T} \vec m)^2$,
in order of decreasing eigenvalue.
The eigenvalue spectrum drops steeply at 2601, as expected, 
when we enter the zero eigenmode regime.
The HRHD spectrum instead flattens off at around $10^{-4}\mu{\mathrm K}^2$.
The high-end power of the HRHD spectrum comes from the multipoles
above $\ell=50$, which are not included in the NCVM. 
The level of the high-multipole plateau is insignificant in comparison with the general noise level;
however, it will blow up in the $\chi^2$ test, which attaches equal weight to all eigenmodes
regardless of their contribution to the total power.

We add a phenomenological regularization constant $\sigma_{\rm reg}^2$ to the diagonal of the NCVM
to account for the extra power.
This is equivalent to assuming an additional white noise component of rms $\sigma_{\rm reg}$.
The value is chosen to match the observed level.
Our procedure is different from the one applied for the PR3 maps \citep{planck2014-a07},
since we are not adding regularization noise to the maps themselves.

The regularization process brings along the danger of making the $\chi^2$ test
results artificially good. In fact, by choosing the constant properly, one can obtain a value of exactly 1
for any matrix-map pair.  To avoid any human selection bias
we apply an automatic procedure where we take the regularization constant to be 
the mean of the HRHD power in the multipole range 2602--3072. 
The values of $\sigma_{\rm reg}$ are listed in Table \ref{table:regnoise},
along with the rms of the HRHD noise map.
The required level of regularization is in all cases very low compared with the actual noise.

\begin{table}
\caption{Regularization constant ($\sigma_{\rm reg}$) and rms of the HRHD map.}
\label{table:regnoise}
\footnotesize
\vskip -7mm
\setbox\tablebox=\vbox{
\newdimen\digitwidth
\setbox0=\hbox{\rm 0}
\digitwidth=\wd0
\catcode`*=\active
\def*{\kern\digitwidth}
\newdimen\signwidth
\setbox0=\hbox{+}
\signwidth=\wd0
\catcode`!=\active
\def!{\kern\signwidth}
\newdimen\decimalwidth
\setbox0=\hbox{.}
\decimalwidth=\wd0
\catcode`@=\active
\def@{\kern\signwidth}
\halign{\hbox to 0.9in{#\leaderfil}\tabskip=2.0em&
    \hfil#\hfil\tabskip=2em&
   \hfil#\hfil\tabskip=0em\cr
\noalign{\doubleline}
\noalign{\vskip 2pt}
\omit&               $\sigma_{\rm reg}$&RMS\cr 
\omit\hfil Noise Map\hfil&[\muK]&[\muK]\cr 
\noalign{\vskip 5pt\hrule\vskip 5pt}
\bf 30\,GHz\cr
\hglue 1em$TT$& 0.0135&  0.4778\cr
\hglue 1em$QQ$& 0.0182& 0.6319\cr
\hglue 1em$UU$& 0.0189& 0.6582\cr
\noalign{\vskip 4pt}
\bf 44\,GHz\cr
\hglue 1em$TT$& 0.0148& 0.4432\cr
\hglue 1em$QQ$& 0.0204&  0.6882\cr
\hglue 1em$UU$& 0.0214& 0.6635\cr
\noalign{\vskip 4pt}
\bf 70\,GHz\cr
\hglue 1em$TT$& 0.0126&  0.3455\cr
\hglue 1em$QQ$& 0.0166&  0.5132\cr
\hglue 1em$UU$& 0.0162&  0.4962\cr
\noalign{\vskip 4pt\hrule\vskip 2pt}
}}
\endPlancktable
\end{table}

The $\chi^2$ results obtained with the regularized pixel--pixel NCVM are given in Table \ref{table:chipix}.
We also report the $TT$ result for cases where we have removed the Galactic region
where signal gradients are large.
We tried  two types of mask: a simple azimuthally symmetric mask
that cuts out 10\,\% or 20\,\% of the sky symmetrically around the Galactic equator, 
and the destriping mask \citep{planck2014-a07} that more accurately follows the structure
of Galactic emission and removes the strong point sources at high latitudes.
We downgrade the destriping mask to $\nside=16$ resolution
and reject all pixels where any of the high-resolution subpixels are missing.
This procedure leads to a lower sky coverage,
especially at 30\,GHz where the 78.7\,\% sky coverage of the destriping mask at $\nside=1024$
translates to 49.5\,\% coverage at $\nside=16$.

The polarization results are already very good for 30 and 70\,GHz without masking. 
However, the mask improves the agreement in $TT$,
indicating that at least part of the discrepancy is due to signal residuals in the HRHD temperature map.

At 44\,GHz there is good agreement in temperature.
In polarization, once again, we observe extra power that is not accounted for in the NCVM.

\begin{table}
\caption{Reduced $\chi^2$ results in pixel space.  We test the $TT$, $QQ$, and $UU$ blocks of the pixel-pixel NCVM against the corresponding half-ring noise estimate, for full sky and for three different masks: a symmetric band around the Galactic plane, removing 10\,\% or 20\,\% of the sky; or the downgraded destriping mask (last row).  The sky coverage in each case is reported in the first column.  The statistical 1$\sigma$ variation is 0.018 for full sky, and 0.019 or 0.020 for the 10\,\% or 20\,\% sky masks, respectively.}
\label{table:chipix}
\footnotesize
\vskip -4mm
\setbox\tablebox=\vbox{
\newdimen\digitwidth
\setbox0=\hbox{\rm 0}
\digitwidth=\wd0
\catcode`*=\active
\def*{\kern\digitwidth}
\newdimen\signwidth
\setbox0=\hbox{+}
\signwidth=\wd0
\catcode`!=\active
\def!{\kern\signwidth}
\newdimen\decimalwidth
\setbox0=\hbox{.}
\decimalwidth=\wd0
\catcode`@=\active
\def@{\kern\signwidth}
\halign{\hbox to 1.1in{#\leaderfil}\tabskip=1.0em&
    \hfil#\hfil\tabskip=2em&
     \hfil#\hfil\tabskip=2em&
   \hfil#\hfil\tabskip=0em\cr
\noalign{\doubleline}
\omit&\multispan3\hfil Reduced $\chi^2$\hfil\cr
\noalign{\vskip -2pt}
\omit&\multispan3\hrulefill\cr
\noalign{\vskip 2pt}
\omit\hfil NCVM Block, $f_{\rm sky}$\hfil&$TT$&$QQ$&$UU$\cr 
\noalign{\vskip 5pt\hrule\vskip 5pt}
 \noalign{\vskip 4pt}
\omit\bf 30\,GHz\hfil\cr
\hglue 0.7em $f_{\rm sky} =100\,\%$& 1.1023& 1.0043&  1.0251\cr
\hglue 3.84em 90\,\%& 1.0751& 1.0041& 1.0029\cr
\hglue 3.84em 80\,\%& 1.0572& 1.0042& 0.9872\cr
\hglue 3.84em 49.5\,\%&  1.0887& 1.0068& 1.0053\cr
\noalign{\vskip 4pt}
\omit\bf 44\,GHz\hfil\cr
\hglue 0.7em $f_{\rm sky} =100\,\%$& 1.0464& 1.1118& 1.0595\cr
\hglue 3.84em 90\,\%& 1.0240&  1.1070& 1.0466\cr
\hglue 3.84em 80\,\%& 1.0120&  1.0938& 1.0470\cr
\hglue 3.84em 71.4\,\%&  1.0108& 1.1198& 1.0499\cr
\noalign{\vskip 4pt}
\omit\bf 70\,GHz\hfil\cr
\hglue 0.7em $f_{\rm sky} =100\,\%$& 1.0362& 1.0026& 1.0047\cr
\hglue 3.84em 90\,\%& 1.0212&  1.0024& 0.9825\cr
\hglue 3.84em 80\,\%& 1.0017&  1.0055& 0.9830\cr
\hglue 3.84em 72.4\,\%& 1.0133& 1.0340& 0.9990\cr
\noalign{\vskip 4pt\hrule\vskip 2pt}
}}
\endPlancktable
\end{table}

The expected 1$\sigma$ statistical variation in $\chi^2$ is 0.018 for full sky, 0.019 for the 10\,\% sky mask,
and 0.020 for the 20\,\% mask. All the 70\,GHz $\chi^2$ test results, as well as 30\,GHz polarization results,
are well within 2$\sigma$ statistical variation. 
This indicates that the residual noise in these map components is well described
by the NCVM, and the maps along with the NCVM can be used for cosmological analysis.
More caution should be taken with the 44\,GHz products.

The basic $\chi^2$ test compresses a complicated noise structure into one number.
It also weights all eigenmodes equally, regardless of how much they contribute to
the noise, and is sensitive to the selection of 
the regularization constant $\sigma_{\rm reg}$.

\begin{figure}
\includegraphics[width=8.8cm]{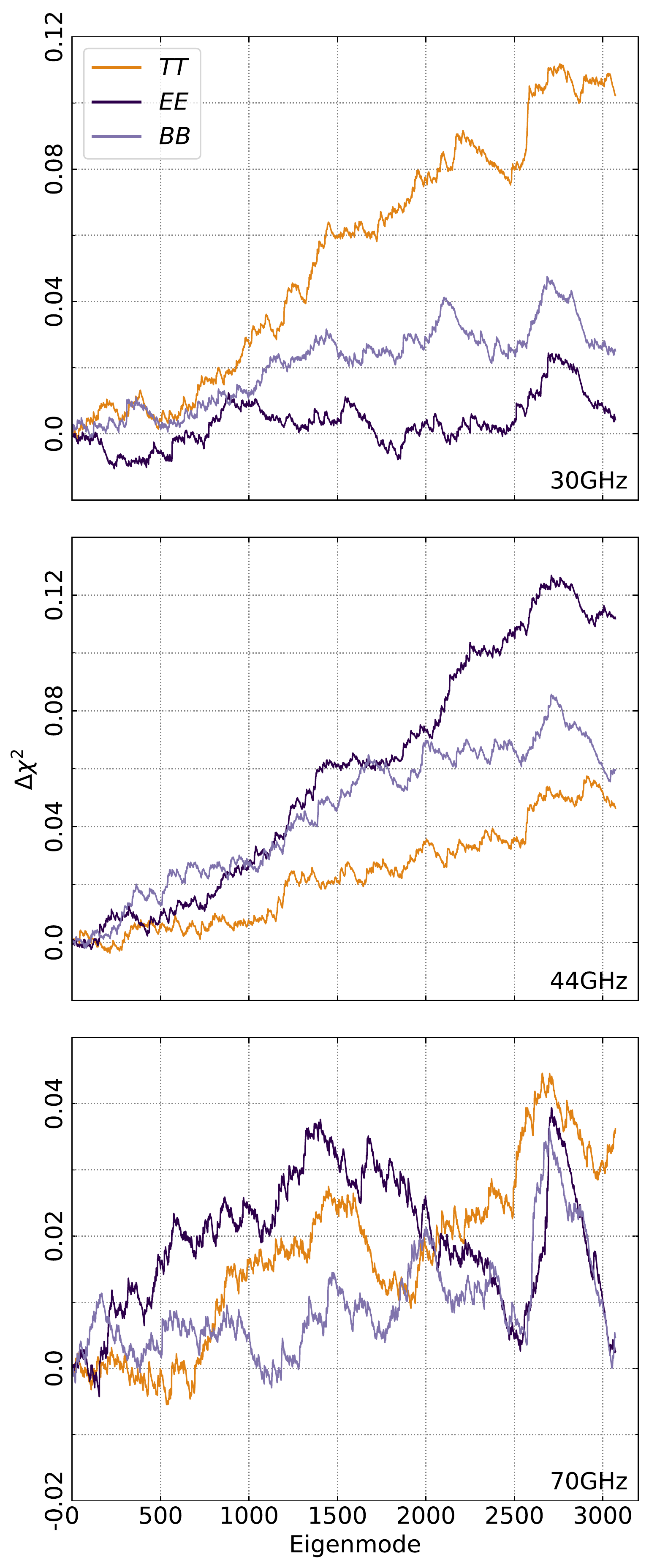}
\caption{Excess $\chi^2$ in pixel space, as it builds up as a function of eigenmode.
The eigenmodes are arranged in order of decreasing eigenvalue,
so that the strongest noise structures are on the left.}
\label{fig:cumuchi}
\end{figure}

We obtain more information by examining how the excess $\chi^2$ builds up.
The noise residuals are by nature extended structures over the sky.  
It is not very meaningful to look at noise amplitudes pixel by pixel.
Instead, we examine them in the eigenmode space.
We introduce the cumulative measure $\Delta\chi^2(n)$, which we calculate as

\begin{equation}
\Delta\chi^2(n) = \frac{1}{3072} \left[ \sum_{j=1}^n\frac{1}{\lambda_j^2} (\vec m_j^{\sf T} \vec e_j)^2 -n \right] . 
\end{equation}
Here, $\vec e_j$ and $\lambda_j$ are the eigenvectors and eigenvalues of the NCVM matrix
in order of decreasing eigenvalue,
and $\vec m_j$ is the decomposition of the noise map under consideration. 
We plot $\Delta\chi^2(n)$ in Fig.~\ref{fig:cumuchi}.
The plot shows how the deviation of $\chi^2$ from unity accumulates
as we add more eigenmodes,
with $\Delta\chi^2(3072)+1$ giving the usual $\chi^2$ test.
The quantity $\Delta\chi^2(n)$ has the convenient property that the statistical variation
remains the same over the plot.
{The cumulative $\Delta\chi^2(n)$
in the high-eigenmode regime is insensitive to
the regularization constant $\sigma_{\rm reg}$,
and thus a more reliable measure of the agreement between NCVM and data
than the single $\chi^2$ value.
}

At both 30 and 70\,GHz, the agreement in the $TT$ block is very good for the first 500 eigenmodes,
after which the excess $\chi^2$ begins to build up gradually.
The strongest noise structures are therefore modelled accurately.
At 44\,GHz, the agreement in $TT$ is good even further out, over 1000 eigenmodes.
The excess $\chi^2$ in 44\,GHz polarization is again evident, more strongly in $EE$ than in $BB$.


\section{Conclusions}
\label{sec:conclusions}

We have produced beam-deconvolved maps from destriped \Planck\ LFI data
using the \artdeco\ deconvolution code.
The \artdeco\ deconvolution approach corrects for all beam-related effects 
and yields a map with a symmetric Gaussian beam.
The chosen smoothing widths are $40'$ (FWHM) for 30\,GHz, $30'$ for 44\,GHz, and $20'$ for 70\,GHz.
These widths are sufficient to remove ringing artefacts and to suppress noise at small scales.

The release includes harmonic coefficients that form the basis of the deconvolved maps.
The interested user can construct deconvolved maps at any desired resolution from those, 
keeping in mind that too narrow a smoothing width will result in ringing around
strong point sources.

We validated the deconvolved maps though analysis of survey difference maps.
Survey difference maps are insensitive to bandpass mismatch,
and reveal the effects of beam mismatch particularly clearly.
We demonstrate that deconvolution removes the beam residuals very well at 44\,GHz.
At 70\,GHz the residuals are small to start with.
With beam residuals cleaned away, 30-GHz survey-difference maps show traces of  
residual signal, which affect the maps at the 10 $\mu$K level.

Deconvolution provides a natural way of extracting the low-resolution component
from \Planck\ data.  We construct low-resolution maps at \healpix\ resolution $\nside=16$,
and corresponding noise covariance matrices (NCVM).  
The NCVM matrices describe the structure of residual noise in the deconvolved maps.

The effects of actual beam asymmetry are small in the
low multipole regime.  It is thus not surprising that the deconvolved low-resolution maps
 resemble the undeconvolved ones.
The value of deconvolution here lies in the natural way that the low-resolution 
regime is separated from the high-resolution regime.
Through deconvolution we avoid the complicated interplay between smoothing 
and pixel-space downgrading (see \cite{planck2014-a07}).

We validate the NCVMs by comparing them against half-ring noise estimates,
both in harmonic space and in pixel space.  
The covariance matrices model the noise well in 30 and 70-GHz polarization maps.
The agreement is less good in temperature, but this is likely due to 
signal residuals in the HRHD estimates rather than the NCVM itself.
At 44\,GHz, we observe a curious asymmetry between $EE$ and $BB$ noise spectra. 
The same asymmetry can also be seen in the undeconvolved spectra.

The deconvolved maps are available through the Planck Legacy Archive.
In any further analysis it should be remembered that, unlike in the case of PR3 maps,
 the polarization component is not corrected for bandpass mismatch,
 and that the residual noise is not well approximated by white noise.

\begin{acknowledgements}
This paper was developed to support the analysis of data from the Planck satellite. The development of Planck has been sup- ported by: ESA; CNES and CNRS/INSU-IN2P3-INP (France); ASI, CNR, and INAF (Italy); NASA and DoE (USA); STFC and UKSA (UK); CSIC, MICINN, JA, and RES (Spain); Tekes, AoF, and CSC (Finland); DLR and MPG (Germany); CSA (Canada); DTU Space (Denmark); SER/SSO (Switzerland); RCN (Norway); SFI (Ireland); FCT/MCTES (Portugal); and PRACE (EU). A description of the Planck Collaboration and a list of its members, including the technical or scientific activities in which they have been involved, can be found at {\tt http://www.sciops.esa.int/index.php?project=planck\&page= Planck\_Collaboration.}
Some of the results in this paper have been derived using the \healpix\
\citep{gorski2005} package.
EK is supported by Academy of Finland grant 295113.
This work was granted access to the HPC resources of CSC made
available within the  Distributed European Computing Initiative by
the PRACE-2IP, receiving funding from the European Community's Seventh
Framework Programme (FP7/2007-2013) under grant agreement RI-283493. 
We thank CSC -- the IT Center for Science Ltd.\ (Finland) -- for computational resources.
We thank the Planck review team - A. Banday, R. Keskitalo, C. Lawrence, D. Maino, D. Scott, and A. Zacchei,
for helpful comments.
\end{acknowledgements}

\bibliographystyle{aa}

\bibliography{Planck_Helsinki_bib}

\end{document}